\documentclass[aps,prl,twocolumn,nofootinbib,longbibliography,superscriptaddress]{revtex4-1}
\usepackage{hyperref}
\usepackage{amsmath}
\usepackage{amssymb}
\usepackage{amsthm}
\usepackage{graphicx}
\usepackage{color}
\newcommand{\abs}[1]{\left\lvert #1 \right\rvert}
\renewcommand{\thanks}[1]{\footnote{#1}}
\newcommand{\doiurl}[2]{{\hypersetup{urlcolor=darkred}\href{http://dx.doi.org/#2}{#1}\hypersetup{urlcolor=blue}}}

\newcommand{\bea}{\begin{eqnarray}}
\newcommand{\eea}{\end{eqnarray}}
\newcommand{\be}{\begin{eqnarray}}
\newcommand{\ee}{\end{eqnarray}}

\def\ie{\begin{equation}\begin{aligned}}
\def\fe{\end{aligned}\end{equation}}
\def\half{{\scriptstyle \frac 12}}

\def\threeh{{\scriptstyle \frac 32}}
\def\fiveh{{\scriptstyle \frac 52}}

\def\ie{\begin{equation}\begin{aligned}}
\def\fe{\end{aligned}\end{equation}}
\def\cG{{\cal A}}

\def\cF{{\cal F}}
\def\cG{{\cal G}}

\def\cI{{\cal I}}

\def\cN{{\cal N}}
\def\cO{{\cal O}}

\def\cQ{{\cal Q}}

\def\cT{{\cal T}}

\def\Zz{{\hat Z}}

\def\Z{{\mathbb Z}}

\def\nn{\nonumber}

\begin{document}

\title{A novel representation of an integrated correlator in $\cN=4$ SYM theory}
\author{Daniele Dorigoni}
\affiliation{ Centre for Particle Theory \& Department of Mathematical Sciences,  Durham University, Durham DH1 3LE, UK}
\author{Michael B. Green}
\affiliation{Department of Applied Mathematics and Theoretical Physics, University of Cambridge, Cambridge CB3 0WA, UK}
\affiliation{School of Physics and Astronomy, Queen Mary University of London, London, E1 4NS, UK}
  \author{Congkao Wen}
\affiliation{School of Physics and Astronomy, Queen Mary University of London, London, E1 4NS, UK}
\begin{abstract}
An integrated correlator of four superconformal stress-tensor primaries of ${\cal N}=4$ supersymmetric $SU(N)$ Yang--Mills theory  (SYM), originally obtained by localisation, is re-expressed as a two-dimensional lattice sum that is manifestly invariant under $SL(2,\Z)$ S-duality. This expression is shown to satisfy a novel Laplace equation in the complex coupling constant $\tau$ that relates the $SU(N)$ integrated correlator to those of the $SU(N+1)$ and $SU(N-1)$ theories.  The lattice sum is shown to precisely reproduce known perturbative and non-perturbative properties of  ${\cal N}=4$ SYM for any finite $N$, as well as  extending previously conjectured properties of the large-$N$ expansion.

\end{abstract}

\maketitle


 $\cN=4$  supersymmetric Yang--Mills ($\cN=4$ SYM) theory  \cite{Brink:1976bc} is a highly non-trivial four-dimensional conformal field theory that is of exceptional interest for a variety of reasons.    It possesses maximal supersymmetry, which enables many of its properties to be determined analytically.  Furthermore, its  relation to string theory in $AdS_5\times S^5$ via the AdS/CFT correspondence provides a model for more general examples of holography. 
 
Of particular significance to this letter is the  analysis of the  integrated correlation function of four superconformal primaries that was formulated {in terms of a $N$-dimensional matrix model in  \cite{Binder:2019jwn}, and further developed in \cite{Chester:2020dja,  Chester:2019pvm,  Chester:2019jas, Chester:2020vyz}.     This integrated  correlator was defined in terms of the partition function of $\cN=2^*$ SYM theory, which is a mass deformation of the superconformal $\cN= 4$ SYM theory with mass parameter $m$. The suitably normalised $\cN=2^*$ $SU(N)$ partition function, on a round $S^4$, $Z_{N}(m, \tau, \bar \tau)$,    
was determined by Pestun  using supersymmetric localisation  \cite{Pestun:2007rz}.  Our notation follows usual conventions where the complex Yang--Mills coupling constant is defined by $\tau = \tau_1+i \tau_2 := \theta/2\pi+ i 4\pi /g_{_{YM}}^2$.

In  \cite{Binder:2019jwn} the integrated correlator  of four superconformal primaries was identified with  the $m\to 0$ limit of four derivatives acting on $\log Z_{N}$ that has the form \footnote{{ The normalisation of the integrated correlator differs from that in \cite{Binder:2019jwn} by a factor of $c/2$ where $c=(N^2-1)/4$ is the central charge of the theory, and we have used the simplified version of the integration measure as given in \cite{Chester:2020dja}.}
}
\ie
\cG_{N}(\tau,\bar\tau)&:= \left.  {1\over 4} \, { \Delta_\tau\partial_m^2 \log Z_{N}}  (m, \tau, \bar{\tau})   \right |_{m=0} \\
& = - {8\over \pi} \int_0^{\infty} dr \int_0^{\pi} d\theta {r^3 \sin^2(\theta) \over U^2} \mathcal{T}_{N}(U,V) \, ,
\label{corrdef}
\fe
 where $\Delta_\tau=4 \tau_2^2\partial_\tau\partial_{\bar\tau}$ is the hyperbolic laplacian and  the cross-ratios  $U, V$ are defined by
\ie
U = {x_{12}^2 x_{34}^2 \over x_{13}^2 x_{24}^2}\, , \qquad V = {x_{14}^2 x_{23}^2 \over x_{13}^2 x_{24}^2} \, ,
\fe
and are related to $r$ and $\theta$ by $U = 1+r^2 -2r \cos(\theta)$ and $V=r^2$.  The function $\mathcal{T}_{N}(U,V)$ is related to the four-point correlator by
\begin{align}
&\langle \cO_2(x_1, Y_1)\dots \cO_2(x_4, Y_4)  \rangle  \\
=&\, {1\over x_{12}^4 x_{34}^4}\left[  \cT_{N,\,\rm free}(U,V; Y_i)+ \mathcal{I}_4(U, V; Y_i)  \, \cT_{N}(U,V) \right]  \, , \nonumber
\end{align}
 where $\cO_2(x_i,Y_i)$ is  a superconformal primary in the ${\bf 20'}$ of the $SU(4)$ R symmetry, which  is encoded in the dependence on the null vectors $Y_i$.  
  $\cT_{N,\,\rm free}(U,V; Y_i)$ is the free field correlator and the pre-factor $\cI_4(U, V; Y_i)$  is determined by superconformal symmetry \cite{Eden:2000bk, Nirschl:2004pa}. So we only focus on the non-trivial part, $\cT_{N}(U,V)$. 
  
  As pointed out in \cite{Binder:2019jwn}, the relation \eqref{corrdef} between the mass derivatives of the localised partition function and the integrated four-point correlator may lead to mixing with long operators, such as the Konishi operator.
Not only do such effects decouple in the large-$N$ strong coupling limit, as argued in \cite{Binder:2019jwn}, but they also do not appear at finite $N$ and finite coupling.\footnote{We thank Shai Chester, Silviu Pufu and Yifan Wang for clarifications on this point.} We will see direct evidence of this statement in our results later in this letter.

 The results in  this letter follow from a reformulation of  $\cG_{N} (\tau,\bar\tau)$, as a  two-dimensional lattice sum that makes manifest many of its properties for all values of $N$ and $\tau$.\footnote{This letter presents our main results but details of the derivation and further results are contained in \cite{DorigoniNew}.}}
These results, which are based on a wealth of evidence concerning the structure of $\cG_{N} (\tau,\bar\tau)$ in various limits, take the form of a conjecture rather than a  mathematical theorem:
    
    \vskip 0.1cm
 {\bf Conjecture}: {\it The integrated correlation function of four superconformal primary operators in the stress tensor multiplet of  $\cN=4$  $SU(N)$ supersymmetric Yang--Mills theory is given by the lattice sum}
\ie
 \cG_{N} (\tau,\bar\tau)  =   \sum_{(m,n)\in\mathbb{Z}^2}  \int_0^\infty e^{- t \pi \frac{|m+n\tau|^2}{\tau_2}}  B_N(t) \, dt\,,
\label{gsun}
\fe
{\it where $B_N(t)$ has the form}
 \ie
 B_N(t)=\frac{\cQ_N(t)}{(t+1)^{2N+1}}\,,
 \label{bndef}
 \fe
 {\it and where  $\cQ_N(t)$ is a polynomial of degree $2N-1$ defined by
\begin{widetext}
\ie
\cQ_N(t) = -\frac{ N (N-1) (1-t^2)^{N+1}}{2(1-t)^2} \left\{ \left(3+  (8N+3t-6) \, t\right ) P_N^{(1,-2)} \left(\frac{1+t^2}{1-t^2}\right)
+ \frac{3t^2-8Nt-3 }  {1+t}   P_N^{(1,-1)}    \left(\frac{1+t^2}{1-t^2}\right)  \right\}\, ,
\fe
\end{widetext}
and $P_N^{(\alpha,\beta)} (z)$ is a Jacobi polynomial. }
\vskip 0.25cm
The following general properties of $B_N(t)$ are of importance in the following,
\ie
B_N(t)= \frac{1}{t} B_N\left(\frac{1}{t}\right)\,,
\label{inverts}
\fe
and 
\ie
\int_0^\infty  B_{N}(t)  dt  = \frac{N(N-1)}{4} \, , \ \  \int_0^\infty B_{N}(t) {1\over \sqrt{t} }  dt  =  0 \,. \label{eq:IntId}
\fe
Using relationships between derivatives of  Jacobi polynomials leads to the recurrence relation
\bea
t\, \frac{d^2}  {dt^2} (t\, B_N(t)) \, 
&&= N(N-1) B_{N+1}(t) - 2(N^2-1) B_N(t) \nn \\
 &&+ N(N+1) B_{N-1}(t)\,.
\label{brecur}
\eea

  The lattice sum (\ref{gsun}) is convergent for $\tau$ in the upper half plane $\tau_2 = \mbox{Im} \tau>0$ and it is manifestly invariant under the $SL(2,\Z)$ transformations 
 \ie
 \tau\to \gamma\cdot \tau = \frac{ a\tau+b}{c\tau +d}\,,\qquad \gamma = \left(\begin{matrix}a & b\\ c&d\end{matrix}\right) \in SL(2,\Z) \, ,
 \fe
 which is in accord with the expectations of Montonen--Olive duality   \cite{Montonen:1977sn, Witten:1978mh,  Osborn:1979tq}.   
 
 An important consequence of \eqref{gsun} together with \eqref{brecur} is that $\cG_N(\tau,\bar\tau)$ satisfies the following corollary:
 
 \vskip 0.1cm
 {\bf Corollary}: {\it  The integrated correlator satisfies a Laplace-difference equation of the form}
 \ie
\left( \Delta_\tau-2 \right)& \cG_{N} (\tau,\bar\tau) = N(N-1) \,\cG_{N+1}(\tau,\bar\tau)  \\
&-2N^2 \cG_N(\tau,\bar\tau)  +N(N+1)\, \cG_{N-1}  (\tau,\bar\tau)\,.
 \label{corollary}
 \fe

 This follows by applying the laplacian $\Delta_\tau$ to \eqref{gsun} and using \eqref{brecur}. Equation \eqref{corollary}  provides powerful constraints on $\cG_N$ that relate the dependence on the coupling $\tau$ and the dependence on $N$ in a manner that will be discussed later.   
For now we note that as $N\to \infty$, assuming $\cG_N$ is a differentiable function of $N$, \eqref{corollary} becomes a Laplace equation in both $\tau$ and $N$, taking the form 
 \ie
 (\Delta_\tau-2)\cG_N (\tau,\bar\tau)\!  \underset{N\to \infty}{ =}  \! (N^2 \partial_N^2  - 2N\partial_N)\,\cG_N (\tau,\bar\tau)  \,,
 \label{lapcon}
 \fe
 where terms of higher order in $1/N$ have been suppressed.

\subsection{The  structure of the integrated correlator}
The $\cN=2^*$ partition function appearing in \eqref{corrdef} has the form \cite{Pestun:2007rz} 
\begin{align}
  Z_{N}(m, \tau, \bar \tau) &\nn=  \int d^N a_i \, \delta(\sum_i a_i) \,  \Big(\prod_{i < j}a_{ij}^2\Big)  e^{- \frac{8 \pi^2}{g_{_{YM}}^2}\sum_k  a_k^2}  \, \\
  & \label{Zdef}\phantom{=}\times  \Zz^{pert}_{N}(m, a_{ij})\,  \abs{\Zz^{inst}_{N}(m, \tau, a_{ij})}^2 \, .
 \end{align}
The perturbative factor in \eqref{Zdef} is given by 
\ie
 \Zz_{N}^{pert}(m, a_{ij}) = H(m)\prod_{i,j} \frac{   H(a_{ij})}{ H(a_{ij}+ m)} \, , 
\fe
 where  $H(z)=e^{-(1+\gamma)z^2}\, G(1+iz)\, G(1-iz)$, and $G(z)$ is a Barnes G-function (and $\gamma$ is the Euler constant).   The factor $|\Zz_{N}^{inst}|^2$ is the product of  Nekrasov partition functions that describes contributions from instantons and anti-instantons localised at the poles of $S^4$ \cite{Nekrasov:2002qd}. 
 
In the following we will consider the Fourier expansion of the integrated correlator, using the notation
 \bea
\cG_{N} (\tau,\bar\tau) := \sum_{k \in \Z} \cG_{N,k} (\tau,\bar\tau)
:=\sum_{k \in \Z}  e^{ 2\pi  i k \tau_1} \cF_{N,|k|}(\tau_2) \,. \ \ \
 \label{modesK}
 \eea
Details of the derivation of the following results are presented in \cite{DorigoniNew}.

 \vskip 0.25cm
\paragraph{(i) Gauge group $SU(2)$.}

When $N=2$, the perturbative contribution arises solely from $\Zz^{pert}_2$ and it is straightforward to show that it can be expressed as an asymptotic series in $g_{_{YM}}^2$ of the form
\bea
\cG^{(i)}_{2 ,0} (y) \sim  \sum_{s=2}^\infty \frac{(2s-1) \Gamma(2s+1) (-1)^s}{2^{2s-1} \Gamma(s-1)} \zeta(2s-1)y^{1-s} \,,  \ \ \ \ \ \
 \label{largey}
 \eea
where $y=\pi\tau_2=4\pi^2/g_{_{YM}}^2$.  This can be resummed by expressing it as a convergent Borel integral
 \ie
\cG_{2,0} (y)=  y \int_0^{\infty}  \frac{ e^{-t} (6t -9t^2 +2t^3)}{2\sinh^2\left(  \sqrt{y t}\right)} dt\,.
\label{intrep}
\fe
In this form it can also be re-expanded at strong-coupling in positive powers of $y$,
\bea
 \cG^{(ii)}_{2 ,0} (y) \sim \frac{1}{2} + \! \sum_{s=2}^\infty(s-1)(2s-1)^2 \Gamma(s+1) \zeta(2s)   \frac{(-y)^s}{\pi^{2s}} \,, \ \ \ \ \  \
\label{smally}
\eea
It will be useful to formally identify the $k=0$ mode of $\cG_{2,0}$ in \eqref{intrep}  with the average of the $y^s$ and  $y^{1-s}$ terms,
\ie
\cG_{2,0}(y) = \frac{1}{2}\left(\cG_{2,0}^{(i)}(y) + \cG_{2,0}^{(ii)}(y)\right) \,.
\label{avzero}
\fe

 The non-zero modes corresponding to instanton and anti-instanton contributions can be extracted from the $|\hat Z^{inst}_2|^2$ factor in  \eqref{Zdef} by extending the analysis in \cite{Chester:2019jas}. This involves a systematic  decomposition of 
 $\Delta_\tau\partial_m^2 \hat Z^{inst}_2(m, \tau, a_{ij}) \big{|}_{m=0}$  in terms of a sum over rectangular Young diagrams with $\hat m$ rows and $n$ columns, where $k= \hat m \, n$ is the instanton number.   The resulting $k$-instanton contribution is
  \begin{align}
&\!\!\! \cG_{2,k} (\tau,\bar\tau)=\\
&\nn \frac{e^{2\pi i k \tau_1}}{2} \!\! \!\!  \sum_{\underset{\hat m n =k}  {\hat{m}\neq 0\,, n\neq0}} \! \!\!\!\!\!\!  & \int_0^\infty    \!  e^{-\pi \tau_2(  \frac{\hat{m}^2}{t} +  n^2 t ) }   \sqrt{\frac{\tau_2}{t}} B_2(t)\, dt\, ,
\label{kinstrev}
\end{align}
with $B_2(t)$ given in \eqref{bndef} for $N=2$.  This integral can  be expanded as an infinite sum of $K$-Bessel functions using the integral representation 
 \ie
\label{kint}
\int_0^\infty e^{-\frac{a^2}{t}- b^2 t} t^{\nu-1}dt = 2\left(\frac{a}{b} \right)^\nu \, K_\nu(2 ab)\, ,
\fe
with $a=\sqrt{\pi \tau_2} \, \hat m $ and $b= \sqrt{\pi/\tau_2} \,n$. 
 
We now recognise that  the total integrated correlator, $\cG_2 =\cG_{2,0}+ \sum_{k\ne 0}\cG_{2,k}$  is an infinite sum  of non-holomorphic Eisenstein series with integer index and with rational coefficients
\ie
\cG_{2}  (\tau,\bar\tau)=  {1 \over 4}+ {1\over 2} \sum_{s=2}^\infty  c^{(2)}_s E(s; \tau,\bar\tau)\, ,
\label{eisensum2}
\fe
 where
  \ie
  \label{cs2def}
 c^{(2)}_s = { (-1)^{s}  \over 2}(s-1) (1-2 s)^2\, \Gamma (s+1) \, . 
 \fe
In making this identification we recall that a non-holomorphic Eisenstein series has a Fourier expansion of the form
\begin{align}
\label{Esdef}
&E(s; \tau,\bar\tau) := \frac{1}{\pi^s} \sum_{(m,n)\neq(0,0)} \frac{\tau_2^s}{|m+ n\tau |^{2s}}\nn\\
&=\frac {2\zeta(2s)}{\pi^s} \tau_2^s  +   \frac{2\sqrt \pi \,\Gamma(s-\half) \zeta(2s-1)}{\pi^s \Gamma(s)}\, \tau_2^{1-s} \\
&+\sum_{k\ne 0} e^{2\pi i k \tau_1} \frac{4}{\Gamma(s)}\,  |k|^{s-\half} \, \sigma_{1-2s}(|k|)
\sqrt{\tau_2}\,K_{s-\half}(2\pi |k|\tau_2) \,. \nn
 \end{align}
 We further note that $E(s,\tau,\bar\tau)$ satisfies the Laplace equation
 \ie
 (\Delta_\tau-s(s-1))\, E(s,\tau,\bar\tau)=0\,.
 \label{lapeisen}
 \fe

 Upon substituting the integral representation
 \ie \label{EsIntdef}
E(s; \tau,\bar\tau) =\sum_{(m,n)\neq(0,0)} \int_0^\infty e^{- t \pi \frac{|m+n\tau|^2}{\tau_2}} \frac{t^{s-1}}{\Gamma(s)} dt \,,
\fe
 into \eqref{eisensum2} it takes the form given in \eqref{gsun} with  $N=2$. 
 
 \vskip 0.25cm
 \paragraph{(ii) Gauge groups $SU(N)$ with $N>2$.}
Here the direct analysis of \eqref{corrdef} is considerably more complicated and is presented  in more detail in  \cite{DorigoniNew}, where the expression for $B_N(t)$ in \eqref{bndef} is motivated.  However, for the purposes of this letter it is more efficient to  use the Laplace-difference equation \eqref{corollary}  to generate the expression for the integrated correlator when $N>2$.
Once we input the boundary conditions $\cG_1=0$ and $\cG_2$ given by \eqref{eisensum2}, the correlators for theories with higher $N$ are generated recursively. They may be expressed as
   \ie
\cG_{N}  (\tau,\bar\tau)=  {N(N-1) \over 8}+ {1\over 2} \sum_{s=2}^\infty  c^{(N)}_s E(s; \tau,\bar\tau)\, ,
\label{eisensum}
\fe
where the coefficients $c_s^{(N)}$ are rational numbers that depend on $N$ and are generated by the expansion of  $B_N(t)$ in the form
  \ie
  \label{bexpand}
 B_N(t) = \sum_{s=2}^{\infty} \frac{c^{(N)}_s }{\Gamma(s)} t^{s-1} \, .
 \fe
 The coefficients $c_s^{(N)}$ can also be determined up to any desired order by substituting the series \eqref{eisensum} into \eqref{corollary} and solving iteratively  in terms of the coefficients $c_s^{(2)}$ given in \eqref{cs2def}.

 \subsection{Properties of the integrated correlator}
 
 The integrated  correlator has interesting behaviour when expanded in various  domains of the parameters, $N$ and $\tau$. We will here discuss three of these domains.
 
 \vskip 0.25cm
   
  \paragraph{(a) Finite $N$, small $\lambda=g_{_{YM}}^2 N$.}
  
  This is the domain of standard Yang--Mills perturbation theory.  The expansion of  the perturbative part of the expression \eqref{gsun}  has the form
\begin{align}
 \label{weak}
&\cG_{N, 0}  (\tau_2) =    (N^2-1) \left[ \frac{3   \, \zeta (3) a   }{2} -\frac{75 \, \zeta (5)a^2}{8} 
+\frac{735 \,\zeta (7) a^3}{16} \right.\nn \\
&\left. -\frac{6615  \,\zeta (9)  \left(1 + \frac{2}{7} N^{-2}\right)  a^4 }  {32}  +\frac{114345 \,  \zeta (11) \left(1+ N^{-2} \right)a^5  }{128 } \right. \nn\\
&\left. -\frac{3864861 \,\zeta(13) \left(1 +  \frac{25}{11}  N^{-2}+ \frac{4}{11}  N^{-4} \right) a^6}{1024} 
+ \mathcal{O}(a^{7}) \right] \, ,
\end{align}
where $a= g_{_{YM}}^2N/ (4 \pi^2)$ and arbitrary $N\ge 2$. When $N=2$, this reduces to \eqref{largey}. 
 Although the perturbative expansion of the unintegrated four-point correlator has a very complicated dependence on the cross ratios $U, V$, the above expression is remarkably simple, consisting of a power series in $a$ with coefficients that are rational multiples of odd Riemann zeta values.
 
  This expansion is in rather impressive agreement with known facts concerning the perturbative expansion of the four-point correlator of superconformal primaries of  $\cN=4$ SYM.   The expressions for the unintegrated correlator up to three loops (up to order $a^3$) are given in \cite{Drummond:2013nda}.  In \cite{DorigoniNew} we have verified the integrals of  the one-loop and two-loop contributions agree with the coefficients proportional to $a$ and to $a^2$ in \eqref{weak}.  In performing these integrals we make use of the all-order results for ladder diagrams \cite{Usyukina:1993ch}.  The one-loop and two-loop contributions are special cases of such ladder diagrams.  However, at higher loops the correlator contains more general diagrams that we have not evaluated. 
  
 A further property that is apparent from the perturbative expansion \eqref{weak} is the dependence on $N$.  We see that up to order $a^3$ the coefficients do not depend on $N$, apart from the overall factor of $(N^2-1)$. This is in accord with known perturbative properties of the correlator, which develop a dependence on $N^{-2}$ at order $a^4$ \cite{Eden:2011we, Fleury:2019ydf}.  From \eqref{weak} we anticipate that an extra power of $N^{-2}$ will appear at every subsequent even power of $a$, which is in agreement with the observations in \cite{Boels:2012ew}.
 
All these precise agreements would be spoilt if there were any additional contribution such as mixing with a three-point correlator involving the Konishi operator.

\vskip 0.25cm

\paragraph{(b)  Large $N$, with fixed $\lambda=g_{_{YM}}^2 N$.}

In this limit instantons are of order $e^{-8\pi^2 k N/\lambda}$ and are therefore suppressed.  The large-$N$ expansion of the   correlator  is 't~Hooft's topological expansion, 
\ie
\cG_N(\tau,\bar\tau)  \sim \sum_{g=0}^\infty N^{2-2g} \,\cG^{(g)}(\lambda)\, , 
\label{eq:genusExp}
\fe
where the leading term is of order $N^2$ and is given by the sum of planar Feynman diagrams in Yang--Mills perturbation theory.
Given our knowledge of the coefficients $c_s^{(N)}$ in \eqref{bexpand} we are able to determine the $\lambda$-dependence of $\cG^{(g)}$ order by order in $N$.   For small $\lambda$ the leading term is  given by the series
\ie
\label{small-lam}
\cG^{(0)} (\lambda) &=&  \sum_{n=1}^{\infty} \frac{4 (-1)^{n+1} \zeta (2 n+1) \Gamma
   \left(n+\frac{3}{2}\right)^2}{  \pi ^{2n+ 1} \Gamma (n) \Gamma (n+3)}  \lambda ^n\,,
  \fe
which converges for $|\lambda| < \pi^2$.  It can be resummed to give   
   \begin{align}
   \label{oldres}
\cG^{(0)} (\lambda)= \lambda \int_0^{\infty} dw \,  w^3 \frac{   _1F_2\left(\frac{5}{2};2,4;-\frac{w^2 \lambda
   }{\pi ^2}\right)}{4 \pi ^2 \, \sinh^2(w)}  \, ,
\end{align}
which is well-defined for $\lambda \geq \pi^2$ and coincides with the result of \cite{Binder:2019jwn} after using an identity that relates $_1F_2$ and Bessel functions $J_{\alpha}$.

However, following an analysis similar to \cite{Arutyunov:2016etw},  it is easy to see that the large-$\lambda$ expansion of \eqref{oldres} is divergent and not Borel summable since the Borel integral, is obstructed by a branch cut along the positive axis.  This signifies that in order to reproduce the exact result \eqref{oldres}  one needs a resurgent  non-perturbative completion $\Delta \cG^{(0)} (\lambda)$, which is determined in  \cite{DorigoniNew}  to be of the form 
  \ie
 \label{G0-resurgence}
\Delta & \cG^{(0)} (\lambda)   =  i\Big(8 \, \mbox{Li}_0(e^{-2\sqrt{\lambda}}) + \frac{18\, \mbox{Li}_1(e^{-2\sqrt{\lambda}})}{\lambda^{1/2}} \\
&+ \frac{117\, \mbox{Li}_2(e^{-2\sqrt{\lambda}})}{4 \lambda} +\frac{489\, \mbox{Li}_3(e^{-2\sqrt{\lambda}})}{16\lambda^{3/2}}+ \cdots \Big) \, .
\fe
 This expression is a sum of  `instantonic'  terms that are non-perturbative in $1/\sqrt \lambda$,  with coefficients  $O (e^{-2\sqrt{\lambda}} )$ that are similar to those found in \cite{Basso:2007wd, Aniceto:2015rua, Dorigoni:2015dha} for the cusp anomalous dimension and other quantities in $\cN=4$ SYM \cite{Basso:2020xts}. 
 Similar arguments lead to non-perturbative completions of $\cG^{(g)}$.  For example, the expression for $\Delta \cG^{(1)}(\lambda)$ is also determined in \cite{DorigoniNew} and takes an analogous form as \eqref{G0-resurgence}.    We believe that   such non-perturbative effects in the large-$\lambda$ expansion  have a holographic interpretation in terms of string world-sheet instantons.

\vskip 0.25cm

\paragraph{(c)  Large $N$, with fixed $g_{_{YM}}^2$.}

This is the large-$N$ limit in which Yang--Mills instantons contribute in a manner that ensures that  $SL(2,\Z)$ S-duality is manifest.  The form of $\cG_N$ can be obtained (as in \cite{DorigoniNew}) by  a  large-$N$ expansion of  $B_N(t)$ (defined in \eqref{bndef}), which is an expansion in half-integer powers of $N$.
It is easy to check that this leads to a solution of  \eqref{corollary} of the form
  \begin{equation}
\!\! \cG_{N} (\tau,\bar\tau)  \sim { N^2 \over 4}+ \sum_{\ell=0}^\infty N^{{\half -\ell}}   \,  \sum_{s=\threeh}^{\ell+\threeh}   \!\! d_\ell^{s}\, E(s; \tau,\bar\tau)   \, , 
\label{eisenlargeN}
\end{equation}
which is a series of Eisenstein series with $s\in \Z+1/2$.   The terms with $s=\ell+\threeh$ satisfy the limiting large-$N$ Laplace equation \eqref{lapcon} but this does not determine  their coefficients, which have to be input from the expansion of $B_N(t)$, giving
  \begin{align}
d_\ell^{{\scriptstyle \ell+ \frac 32}} &\label{eq:LeadingEisd} = \frac{(\ell+1) \Gamma\Big( \ell-\frac{1}{2}\Big)\Gamma\Big(\ell+\frac{3}{2}\Big)\Gamma\Big(\ell+\frac{5}{2}\Big)}{2^{2\ell+2} \pi^{3/2}\, \ell!} \,. 
  \end{align}
 Once $d_\ell^{{\ell+  \scriptstyle \frac 32}}$ is input the Laplace-difference equation determines the rest of the solution.  
This  reproduces and extends the results of  \cite{Chester:2019jas}, where the first few coefficients were obtained. 
  For example, terms with $s=\ell-\half >0$ and $s=\ell-\fiveh>0$ are given by 
\bea
d_\ell^{\ell {- \scriptstyle \frac 12}}  &=& -\frac{(\ell-1)^2(2\ell+9) \Gamma\left(\ell-\frac{1}{2}\right)\Gamma\left( \ell+\frac{1}{2}\right)^2}{3\,2^{2\ell+3} \pi^{3/2}\, \ell!}  \,,\nn\\
d_\ell^{\ell- {\scriptstyle \frac 52}} \label{eq:subsubLeadingEisd}&=&\, \frac{ (\ell-3)^2 (20\ell^2 + 48\ell-293)  \Gamma
   \left(\ell-\frac{5}{2}\right)  }{45 \, 2^{2 \ell+5} \pi ^{3/2} \Gamma
   (\ell)} \cr
 && \quad \Gamma \left(\ell-\threeh \right) \Gamma \left(\ell+ \threeh \right) \,.  
\eea

 Finally, we believe that the considerations of this letter generalise to a second integrated correlator that was considered in~\cite{Chester:2020dja} and  further explored in  \cite{Chester:2020vyz}. This   is obtained from the $\cN=2^*$ partition function by applying four derivatives with respect to mass,
$\cG'_N(\tau,\bar\tau) := \left.  { \partial_m^4 \log Z_N}(m, \tau, \bar{\tau})    \right |_{m=0}$,
which again generates a supersymmetric  integrated correlator of four superconformal primaries, but with a different integration measure.

\section*{acknowledgments}
{\small 
 We would like to thank  Shai Chester, Lance Dixon,  Paul Heslop, Silviu Pufu, Yifan Wang,  and Gang Yang,  for useful conversations and comments.
 DD would like to thank the Albert Einstein Institute for the hospitality and support during the writing of this paper.
 MBG has been partially supported by STFC consolidated grant ST/L000385/1.   CW is supported by a Royal Society University Research Fellowship No. UF160350.

\end{document}